\documentclass[aps,prl,twocolumn, superscriptaddress]{revtex4-2}

\usepackage{graphicx}
\usepackage{amssymb}
\usepackage{amsmath}
\usepackage{xspace}
\usepackage{longtable}
\usepackage{soul}
\usepackage{xcolor}
\usepackage{epstopdf}
\usepackage{array}
\usepackage{tabularx}
\usepackage{physics}
\usepackage{float}
\usepackage{comment}
\usepackage{dsfont}
\usepackage[utf8]{inputenc}
\usepackage{siunitx}

\usepackage[colorlinks = true,linkcolor = blue,urlcolor  = blue,citecolor = blue,anchorcolor = blue]{hyperref}

\newcommand{\HoW}{HoW$_{10}$\xspace}

\newcommand{\HoWF}{$[$Ho(W$_5$O$_{18}$)$_2]^{9-}$\xspace}
\newcommand{\HoWEn}{Na$_{9}$[Y$_{1-x}$Ho$_x$(W$_5$O$_{18}$)$_2$]$\cdot$35H$_2$O\xspace}

\begin{document}
\title{Electrical two-qubit gates within a pair of clock-qubit magnetic molecules}
\author{Aman Ullah }
\thanks{These authors contributed equally to this work}
\affiliation{Instituto de Ciencia Molecular (ICMol), Universitat de Val\`encia, Paterna, Spain}

\author{Ziqi Hu }
\thanks{These authors contributed equally to this work}
\affiliation{Instituto de Ciencia Molecular (ICMol), Universitat de Val\`encia, Paterna, Spain}

\author{Jesús Cerdá}
\affiliation{Instituto de Ciencia Molecular (ICMol), Universitat de Val\`encia, Paterna, Spain}
\author{Juan Aragó}
\email{juan.arago@uv.es}
\affiliation{Instituto de Ciencia Molecular (ICMol), Universitat de Val\`encia, Paterna, Spain}
\author{Alejandro Gaita-Ariño}
\email{alejandro.gaita@uv.es}
\affiliation{Instituto de Ciencia Molecular (ICMol), Universitat de Val\`encia, Paterna, Spain}

\date{\today}

\begin{abstract}

Enhanced coherence in \HoW molecular spin qubits has been demonstrated by use of Clock Transitions (CTs).
More recently it was shown that, while operating at the CTs, it was possible to use an electrical field to selectively address \HoW molecules pointing in a given direction, within a crystal that contains two kinds of identical but inversion-related molecules.
Herein we theoretically explore the possibility of employing the electric field to effect entangling two-qubit quantum gates among two neighbouring CT-protected \HoW qubits within a diluted crystal.
We estimate the thermal evolution of $T_1$, $T_2$, find that CTs are also optimal operating points from the point of view of phonons, and
lay out how
to combine a sequence of microwave and electric field pulses to achieve coherent control within a 2-qubit operating space that is protected both from spin-bath and from phonon-bath decoherence. Finally, we found a highly protected 1-qubit subspace resulting from the interaction between two clock molecules.

\end{abstract}

\maketitle




Electrical control of spins at the nanoscale offers significant architectural advantages in spintronics, because electric fields ($E$-fields) can be confined over shorter length scales than magnetic fields ($B$-fields) \cite{kane1998silicon, trif2008spin, laucht2015electrically, tosi2017silicon, asaad2020coherent}. In the context of qubits, the use of $E$-fields has been already suggested as a strategy to generate entangling two-qubit gates, 
either passing current pulses through molecules in single-molecule spintronics, or by applying $E$-fields in a P@Si crystal\cite{Loss1998,Lehmann2007,godfrin2017operating,madzik2021conditional,savytskyy2022electrically}.


Magnetic molecules are considered as an ideal platform in this line since their spin Hamiltonian can be tailored by chemical design\cite{gaita2019molecular}. Constructing two-qubit gates has been already explored theoretically based on the magnetic interaction between assymetric centers in molecular nanomagnets\cite{luis2011molecular,aguila2014heterodimetallic} or supramolecular dimers\cite{collett2020constructing,ferrando2016modular,timco2009engineering}. 
However, implementing individual qubit addressing and two-qubit logical gates as independent physical operations is challenging in practice.
Another challenge for molecular spin qubits is that they exhibit fragile quantum coherence owing to the inevitable interaction with the environment (spin and phonon baths). A significant enhancement has been recently achieved via clock-transitions (CTs), which protect from magnetic noise\cite{shiddiq2016enhancing,kundu20229,rubin2021chemical}. Nevertheless, at the CT, the targeted molecular qubit would behave diamagnetically, thus inhibiting inter-qubit communication and impeding the implementation of a two-qubit quantum gate. We will explore herein how these limitations may be overcome by slightly moving away from the CT but preserving some coherent protection and, at the same time, the use of a directional $E$-field to electronically tune the transition frequencies of the two interacting qubit molecules.

\begin{figure}[tbh!]
\centering
\includegraphics[width=3.4in]{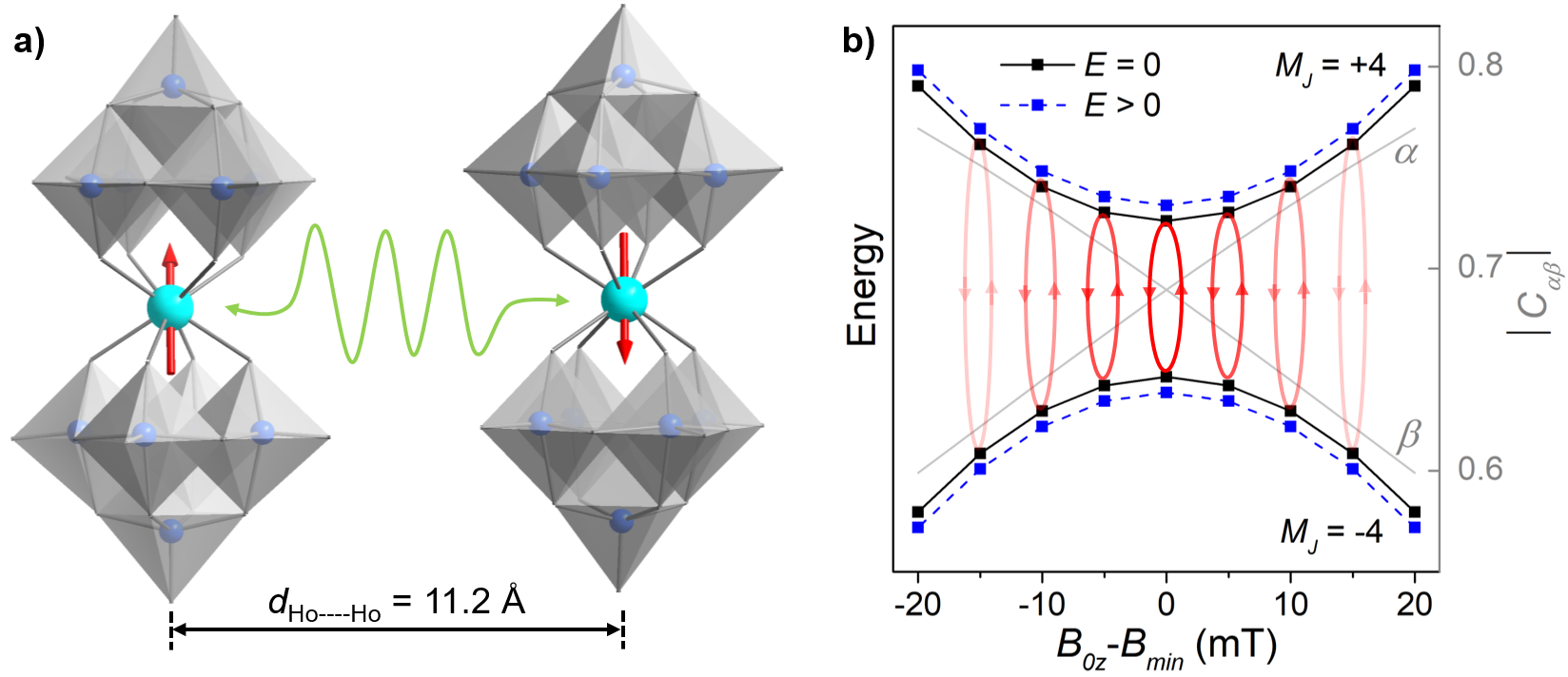}\\
\caption{a) Nearest inversion-related \HoW pair within the crystal, illustrating their dipolar interaction. Their magnetic axes point in opposite directions. b) Calculated energies (left axis) and admixing of wavefunction for $\left|M_{J} = \pm 4\right>$ of \HoW with corresponding $\alpha$ and $\beta$ coefficients (right axis). The first CT is located at $B_{min}$ = 24 mT. The calculated CT frequency is $\simeq 11$ GHz ($\simeq 9.1$ GHz in experiment).} 
\label{Fig:WaveFn}
\end{figure}

The polyoxometalate molecular anion \HoWF (abbreviated to \HoW) with crystal structure \HoWEn ($x=1\%$) is a prime example of CT spin qubit\cite{shiddiq2016enhancing}. The crystal unit cell contains two inversion-symmetry related \HoW anions, slightly distorted ($D_{4d}$ symmetry) along their $C_4$ axes (Fig. \ref{Fig:WaveFn}a). The magnetic levels of \HoW can be described by a Hamiltonian including crystal-field,  hyperfine and Zeeman interactions (Eq. S7). The low-energy region of interest comprises 16 electro-nuclear spin levels corresponding to a tunneling-splitting electronic spin doublet $M_J=\pm4$ and a nuclear spin $I=7/2$ (Fig. S1). Each CT corresponds to an anticrossing between the 8th and the 9th levels (see the Supplementary Material (SM), section S2 for additional details). 

A recent experimental study showed that one can achieve coherent control over the spin of \HoW molecules by using an $E$-field pulse to manipulate the CT frequency (Fig. \ref{Fig:WaveFn}b)\cite{liu2021quantum}. This is realized in practice due to a strong spin-electric coupling (SEC) which arises from intrinsic symmetry breaking, a soft and electrically polarizable environment of the spin carriers, and a spin spectrum that is highly sensitive to distortions. 
This allows to selectively address the spins of identical \HoW molecules pointing at different directions.

In this Letter, we theoretically demonstrate the possibility of employing $E$-field pulses 
to effect entangling two-qubit quantum gates among two neighbouring CT-protected molecular spin qubits of \HoW within a diluted crystal. This involves finding a $B$-field constituting a compromise between keeping some of the protection from magnetic noise and allowing the necessary amount of inter-qubit communication. We start with a theoretical estimate of the effect of temperature and $B$-field on the longitudinal ($T_1$) and transverse ($T_2$) relaxation times of a \HoW single qubit. Later, we address the qubit-qubit interaction. With the whole Hamiltonian, we indicate the conditions and procedure to implement arbitrary 2-qubit manipulations in this system.

\textcolor{black}{\emph{Single-qubit relaxation}}. Prior to addressing the two-qubit system (i.e., the two inversion-symmetry related \HoW units of the crystal cell), let us analyze the relaxation dynamics of an individual \HoW molecule. Note that, despite the CT protection, \HoW presents relatively short coherence times $T_2 (5\textrm{K})\simeq 8 \mu\textrm{s}$. We are interested in the role of temperature because of the lack of such information both in experiment and in theory, since previous works have focused on the role of the spin bath for this compound \cite{escalera2019decoherence, chen2021electron}. 

The dynamics of the entire system (electronic spins and phonons) can be described by the time evolution of the density operator, $\hat{\rho}(t)$.
\begin{equation}\label{eq:R1}
	\dot{\hat{\rho}}(t)  = -\frac{\dot{\iota}}{\hbar} \left[\hat{H}, \hat{\rho}(t)  \right]
\end{equation}
where $\hat{H}= \hat{H}_S+ \hat{H}_{ph} + \hat{H}_{S-ph}$ is the total Hamiltonian describing the electronic spins ($\hat{H}_S$), the phonon bath ($\hat{H}_{ph}$) and their spin-bath interaction ($\hat{H}_{S-ph}$), respectively. When phonon dynamics is much faster than the spin relaxation (as it is the case here), the Born-Markov approximation is safely invoked integrating out the phonon component from the density matrix and making the problem purely electronic in the presence of a phonon bath \cite{breuer2002theory}. In this regime, the dynamics of the electronic spin states can be described by the reduced spin density operator ($\hat{\rho}^{S}$) within the Redfield equation in the eigenvector representation of the spin Hamiltonian $\hat{H}_S$ \cite{lunghi2020limit,briganti2021complete,lunghi2021towards,gu2020origins}, 
\begin{equation}\label{eq:R2}
		\dot{\rho}_{ab}^{S}(t) = -\dot{\iota}\omega_{ab}\rho_{ab}-\sum_{cd}R_{ab,cd}\rho_{cd}^{S}(t)
\end{equation}
where $\omega_{ab}=(E_{a}-E_{b})/\hbar$, and $E_{a}$ and $E_{b}$ are the corresponding eigenvalues. $R_{ab,cd}$ is the full Redfield tensor which accounts for the system relaxation due to the interaction with the thermal bath. To evaluate the Redfield tensor, we need a spectral function for the bath, which is taken from \cite{lunghi2017role}, and the spin-phonon couplings, which are estimated from \textit{ab initio} calculations (see SM section S3 for further details).

\begin{figure}[tbh!]
\centering
\includegraphics[width=3.5in, height=3.3in ]{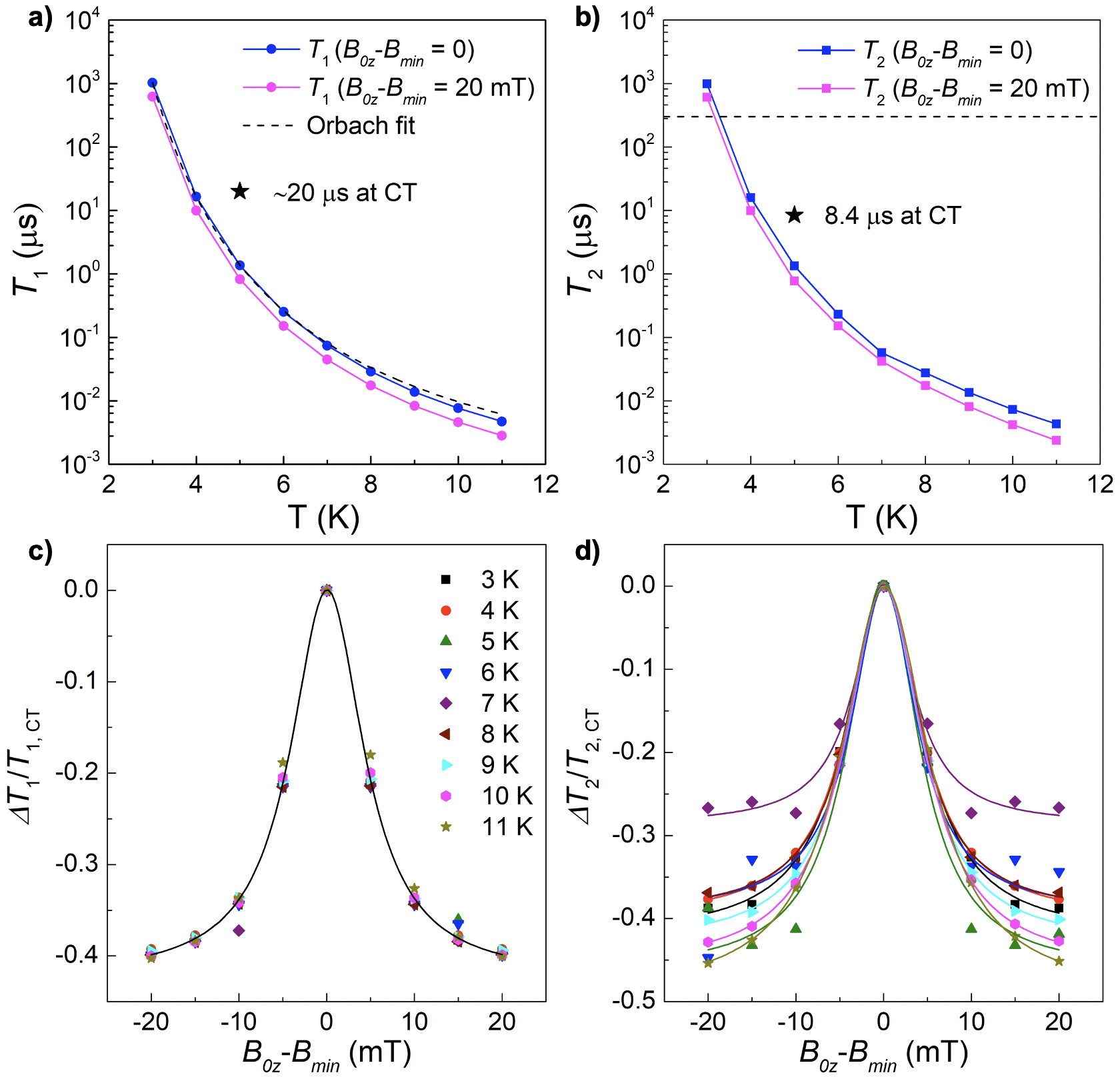}\\
\caption{a) Calculated $T_1$ and b) $T_2$ as a function of temperature at the CT and at 20 mT away from the CT. The dashed line in b) indicates the $T_2$ limit of 300 $\mu s$ estimated from nuclear spin bath in Ref.\cite{escalera2019decoherence}. c) $T_1$ and d) $T_2$ divergences around the CT at different temperatures. The curves serve as a guide for the eye.}
\label{fig:Relax}
\end{figure}

We then solve the master equation (Eq. \ref{eq:R2}) in time at different temperatures. $T_1$ and $T_2$ are thus extracted by fitting the exponential decays of magnetization at any temperature (see SM). The temperature evolution of $T_1$ 
predicted for \HoW (Fig. \ref{fig:Relax}a) reveals an exponential $T_1$-T dependence over the temperature range of 3-11 K. An Arrhenius fit of this Orbach process, $T_1$ $\propto$ $exp(U{_{eff}}/k_BT)$, gives us an effective energy barrier $U_{eff}$ of 34.5 cm$^{-1}$, which is virtually identical to the half of the energy of the lowest-frequency molecular vibration of \HoW (68.4 cm$^{-1}$) \cite{blockmon2021spectroscopic}. This relation is in line with the interpretation of the under-barrier relaxation in molecular nanomagnets by Lunghi et al.\cite{lunghi2017role} and indicates that the longitudinal relaxation of \HoW within the ground doublet is assisted by the lowest-frequency phonon mode at low temperatures. Our analysis also shows that $T_2$s are in the same order as for $T_1$s, both following the similar temperature trend (Fig. \ref{fig:Relax}b). This behavior is in accordance with the observed drastic $T_2$ decrease upon heating\cite{shiddiq2016enhancing}, demonstrating that $T_2$ is limited by $T_1$ which is mainly governed by the spin-phonon coupling with the lowest-frequency phonon mode. We notice that our calculation underestimates $T_1$ and $T_2$ compared with the experimental estimates determined at the CTs. This is likely due to the overestimation of spin-phonon couplings computed within a single \HoW model in gas phase. 


To gain an insight into the relaxation behavior in the vicinity of CTs, relaxation times are further analyzed at different $B$-fields and different temperatures. Fig. \ref{fig:Relax}c and \ref{fig:Relax}d illustrate $T_1$,$T_2$ divergences with the longest relaxation times observed at the CT. 
This protection against vibrational decoherence at the CT coincides with the well-known $T_2$ protection from magnetic noise. The latter originates in the fact that the Ho spin possesses a vanishing magnetic moment resulted from mixing between $\ket{M_J=+4}$ and $\ket{M_J=-4}$ (Fig. \ref{Fig:WaveFn}b). When moving away from the CT, such mixing is broken by the Zeeman effect and thereby the dipolar decoherence is activated. Since we did not include dipolar decoherence in our model, it is not surprising that the calculated $T_2$ drop ($\sim$ 40\%) at 10 mT away from the CT is less intense than the sharp $T_2$ divergence determined in experiments.




\begin{figure*}[tbh!]
\includegraphics[width=\textwidth,height=9cm]{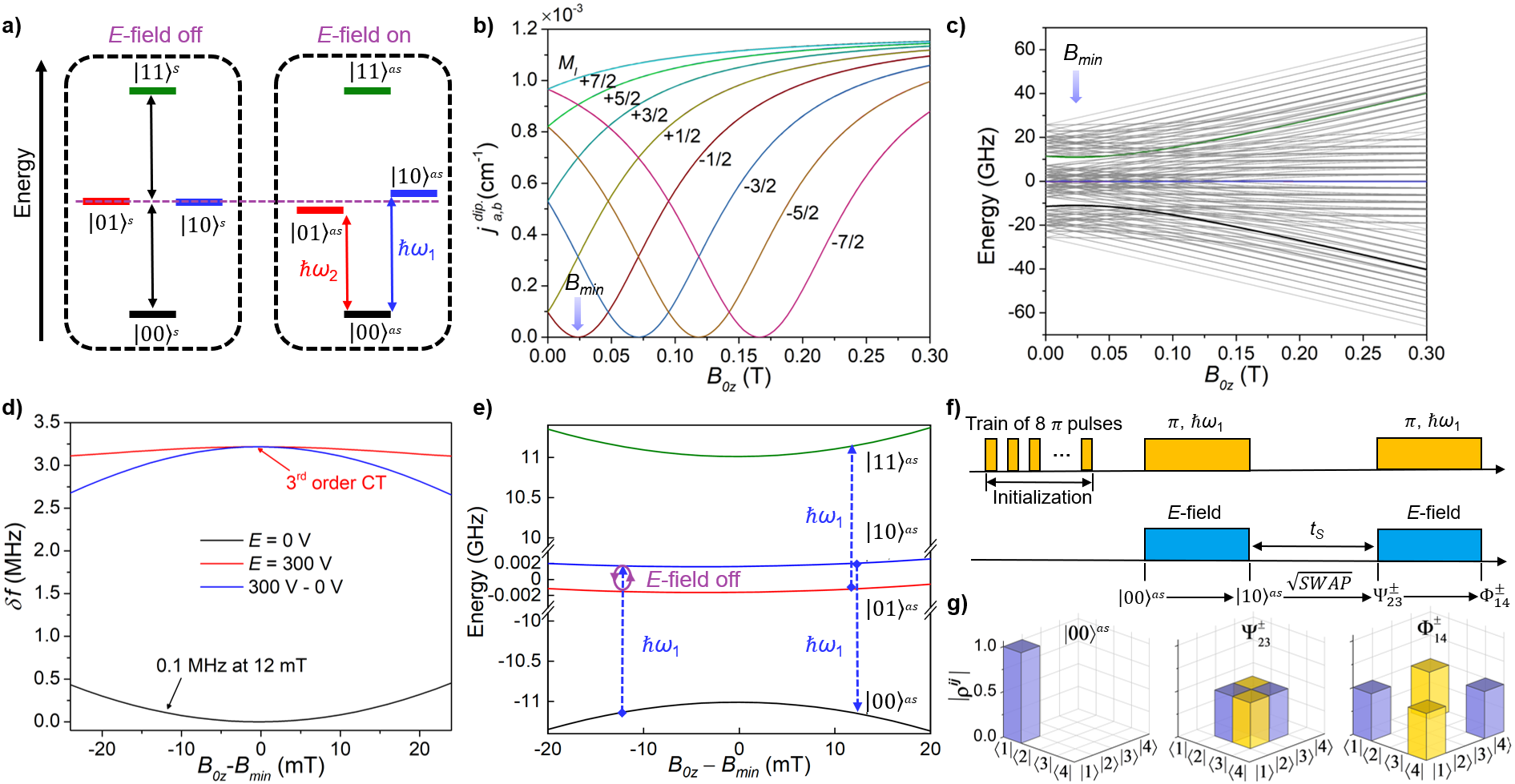}
\caption{a) Scheme of the qubit states in absence and presence of an $E$-field. b) Calculated dipolar exchange ($j_{a,b}^{dip.}$) of the 16 spin levels of a \HoW molecule interacting with its neighboring molecule. The $M_J=\pm4$ levels with the same $M_I$ are overlapped. c) Energy levels of the coupled \HoW dimer in the presence of an $E$-field (300 V) with 2 mm interelectrode distance. The four operating levels are highlighted. The two middle levels (red and blue) are close to each other. d) $\delta f$ as a function of $B$-field with (red) and without (black) applied $E$-field. The SEC effect is indicated by their difference (blue). 
e) The operating space around the first CT and the generation of the entangled states $\Psi_{23}^{\pm}$ and $\Phi_{14}^{\pm}$ at 12 mT (see text). 
f) Pulse sequence and g) related initial and Bell states.}
\label{fig:levels}
\end{figure*}

\textcolor{black}{\emph{Two-qubit gates}}. To explore the possibility of coherent control over a pair of clock qubits (Fig. \ref{fig:levels}a), including generating entangled states, we considered two nearest \HoW qubits as indicated in Fig. \ref{Fig:WaveFn}a, which are interacting by dipolar interactions as a weak perturbation. A technical difficulty of operating on molecular spin qubit pairs in a disordered, diluted system is that isolated molecules with similar transition energies will be far more abundant than molecule pairs. We will outline two independent ways of addressing this problem, based on initialization and on two-qubit gates.

For simplicity, instead of applying opposite biases due to the $E$-field, we apply the $E$-field effect on just one of the molecules. The microscopic description in terms of an effective Hamiltonian of this system is as follows:
\begin{equation}\label{eq:D1}
	\hat{H}^{tot.}  = \hat{H}_s^{a} + \hat{H}_s^{b}(E) + \hat{H}^{ex} 
\end{equation}
\begin{equation}
	\hat{H}^{tot.}  = \hat{H}_s^{a}\otimes\mathds{I}_{b} + \mathds{I}_{a}\otimes \hat{H}_s^{b} + j_{a,b}^{dip.}(B)J_{a}\otimes\mathds{I}_{b}\cdot\mathds{I}_{a}\otimes J_{b}
\end{equation}
where $\hat{H}_s^{a}$ and $\hat{H}_s^{b}(E)$ are the spin (crystal field + hyperfine + Zeeman) Hamiltonians for \HoW sites \textit{a} and electrically-tuned \textit{b}. The crystal field parameters (CFPs) $B_k^q(E)$ are determined by establishing a relation between the given $E$-field and spin energy levels (see methodology in the SM and ref. \cite{liu2021quantum}). $\hat{H}^{ex}$ denotes the interacting Hamiltonian, which accounts for the dipolar interaction ($j_{a,b}^{dip.}$) between the two sites \textit{a} and \textit{b}. This dipolar interaction is dependent of $B$-field for a given two-site distance and orientations (see SM section S5), and vanishes at the CTs where the molecule is effectively diamagnetic (Fig. \ref{fig:levels}b).

Without loss of generality, and since our Hamiltonian does not include terms that are extradiagonal in the nuclear spin, we consider only states involved in the first CT at $B_{min}$ = 24 mT. Within the two dipolarly-coupled molecules, the 16 levels of the individual molecules are combined into a 256 manifold (Fig. \ref{fig:levels}b). Our operating two-qubit space is defined as the 4 levels resulting from the weak dipolar coupling between $\ket{M_J=\pm 4,M_I=-1/2}^{(a)}$ and $\ket{M_J=\pm 4,M_I=-1/2}^{(b)}$ states for sites \textit{a} to \textit{b}, respectively. We label these four electro-nuclear spin states as $\ket{00}$, $\ket{01}$, $\ket{10}$, and $\ket{11}$, independently of their physical nature, which will depend on the applied $B$- or $E$-field. 

In the symmetric scenario (at the CT and zero $E$-field), one would define all four states as linear combination of the single-molecule CT states. Specifically, $\ket{00}^s$ and $\ket{11}^s$, which are energetically well-separated, correspond to the ``double ground'' and ``double excited'' CT states, respectively, and $\ket{01}^s$ and $\ket{10}^s$ are degenerate in energy (Fig. \ref{fig:levels}a, left). Since the two molecules are identical and their dipolar coupling is zero at the CT region, the states of the two \HoW molecules in $\ket{01}^s$ and $\ket{10}^s$ are separable and independent. That electronic situation is better understood not actually as a two-qubit system, but rather as two single-qubit systems that happen to be close to each other and function as two copies of the same qubit.

To generate asymmetric states from two otherwise identical molecules, we need to apply an $E$-field, which makes $\delta f=E(\ket{10}^{as})- E(\ket{01}^{as})\neq 0$ \cite{liu2021quantum} (Fig. \ref{fig:levels}a, right). 
The novelty in the present case is the presence of dipolar coupling among the close qubit neighbours which is activated by Zeeman effect. The two qubits thus no longer correspond directly to single-molecule states, but this does not disturb single-qubit adressability, since transition frequencies are still distinct.
The mixing is evidenced from the wavefunction composition of the coupled \HoW pair (Table S5-8). This dipolar-coupling character, that symmetrizes the wavefunction, is competing with the $E$-field that makes the two molecules inequivalent. Figure \ref{fig:levels}d illustrates $\delta f$ as a function of the deviation from the CT. A high $B$-field is not actually required to achieve this inequivalence, indeed around $B=12$ mT the exchange becomes non-negligible ($\delta f$ = 0.1 MHz).

This exchange framework is able to effect two-qubit gates, but first we briefly address the issue of initialization since our starting state would be the $\ket{-4,-1/2}^{(a)}\otimes\ket{-4,-1/2}^{(b)}$ (Fig. \ref{fig:levels}b and \ref{fig:levels}d, black). One can apply a sequence of pulses to transfer the population from the ground state of the bimolecular Hamiltonian to the ground state of our operating space. We estimated a possible initialization sequence by analyzing the wavefunction within a hyperfine basis ($\ket{M_I^{a},M_I^{b}}$, SM section S6). Note that the dipolar interactions are strong enough that this initialization step is able to distinguish between isolated molecules and pairs of neighbouring \HoW qubits within the diluted crystal. 


 Let us employ a specific case for illustration of single- and double-qubit operations, namely the generation of a Bell state involving $\ket{00}^{as}$ and $\ket{11}^{as}$. Once the initial state $\ket{00}^{as}$ is prepared, the $E$-field is turned on and a microwave $\pi$-pulse is applied to promote the $\ket{00}^{as}\to \ket{10}^{as}$ transition (Fig. \ref{fig:levels}e and \ref{fig:levels}f). The eigenstates after switching off the $E$-field are $\ket{10}^{s}$ and $\ket{01}^{s}$, thus Rabi-like oscillations start between $\ket{10}^{as}$ and $\ket{01}^{as}$. These coherent oscillations constitute a two-qubit $SWAP$ gate by switching $\ket{10}^{as}\to \ket{01}^{as}$. More exactly, any desired rotation between these two states can be achieved by choosing the time of the operation, including the notable $\sqrt{SWAP}$ that together with single-qubit rotations forms a universal gate set. The $\sqrt{SWAP}$ gate operation would result in entangled states that can be read-out in Bell states as $\Psi_{23}^{\pm}=\frac{1}{\sqrt{2}}(\ket{01}^{as}\pm\ket{10}^{as})$. To generate the desired Bell state, applying a $\pi$-pulse to $\Psi_{23}^{\pm}$ enables population transfer from $\ket{10}^{as}\to \ket{00}^{as}$ and $\ket{01}^{as}\to \ket{11}^{as}$ simultaneously because of their indistinguishable energy gaps, i.e. $\Phi_{14}^{\pm}=\frac{1}{\sqrt{2}}(\ket{00}^{as}\pm\ket{11}^{as})$ (Fig. \ref{fig:levels}g). In other words, the transition frequency corresponding to each qubit is independent on the state of the two-qubit system. This unambiguous correspondence between logical operation and physical operation constitutes a key difference between our proposal and previous approaches\cite{luis2011molecular,aguila2014heterodimetallic,timco2009engineering,jenkins2017coherent,luis2020dissymmetric,ferrando2016modular,collett2020constructing}.

Let us give some estimates on practical details. The typical times for two-qubit gate operations will be given by the inverse of the interaction energy between the two molecular spins. With $\delta f=0.1$ MHz at 12 mT as discussed above, a half rotation ($\sqrt{SWAP}$) would take in the order of 5 $\mu$s. The duration of microwave $\pi$-pulses is determined as 800 ns to selectively excite qubits with narrow frequency around 3 MHz in the presence of the $E$-field\cite{liu2021quantum}. Thus, the overall time needed for the pulse sequence illustrated in Fig. \ref{fig:levels}f is comparable to the $T_2$ value at 5 K, but conveniently below the estimated $T_2$ times at 3 or even 4 K, meaning any cooling below 5 K would suffice.

The pair of states $\ket{10}^{as}$ and $\ket{01}^{as}$ merit a separate discussion. We have employed them here as two of the four states in a two-qubit system, but they also could be employed as an exceptionally protected single qubit ($3^{rd}$-order CT, Fig. S11), with the other states being auxiliary. The idea that inter-qubit dipolar interactions and spin–phonon interactions need to be suppressed, and the proposal to do that by having antiferromagnetically ordered spin qubits via chemical design has been around for some time now\cite{stamp2009spin}.
A similar strategy has been achieved experimentally in so-called flip-flop qubits\cite{tosi2017silicon}. One can appreciate the unusual protection against magnetic noise in the energy differences in Fig. \ref{fig:levels}d, and in the energy levels themselves in Figs. S8 and S9.


As mentioned above, a key difficulty from working with pairs of magnetic entities within a diluted crystal is getting past the signal from the monomers, which will be statistically much more abundant. To supress single-qubit signal one needs to design pulse sequences, as in the example above, where all single-qubit operations add up to full $2\pi$ single-qubit rotations, so all the single qubits will be back to the ground state and not contribute to the detected signal. For qubit pairs, which experience two-qubit rotations, the same pulse sequence results in non-trivial operations. The extension of the same strategy to other quantum circuits is discussed in SM section S7, where other challenges for this scheme that may arise from the low symmetry of the crystal structure are discussed, together with possible strategies to address them.

\textcolor{black}{\emph{Discussion and outlook}}. Our work presents both a general methodology to study molecules and their viability as $E$-field controlled CT qubits and a promising strategy towards the use of \HoW molecular spin qubits in constructing two-qubit quantum gates. The inversion-symmetric \HoW pair in a diluted crystal offers several unique features including the robust coherence close to the CTs, the strongest spin-electric response and the resultant switchable operating space. Finally, note that the existence of a highly protected subspace might be exploited for further schemes in a way that this two-qubit system can be considered as two physical qubits able to temporarily store a single logical qubit, as an approach to built-in quantum error protection.

\begin{acknowledgments}
This work is supported by the European Commission (FET-OPEN project FATMOLS (No  862893) ); the Spanish MICINN (grant
CTQ2017-89993 and PGC2018-099568-B-I00 co-financed by FEDER, grant MAT2017-89528 and the Unit of
excellence “María de Maeztu” CEX2019-000919-M); and the Generalitat
Valenciana (CIDEGENT/2021/018 and PROMETEO/2019/066). J.A. is indebted to the MICINN for his "Ramón y Cajal" fellowship (RyC-2017-23500). 
We thank J. Liu for his insightful comments.
\end{acknowledgments}

\bibliographystyle{apsrev4-2}
\bibliography{Reference}

\end{document}